\documentclass[12pt]{article}

\newcommand{\x}{arXiv:}
\newcommand{\m}{\mathrm}
\newcommand{\be}{\begin{equation}}
\newcommand{\ee}{\end{equation}}
\newcommand{\ba}{\begin{eqnarray}}
\newcommand{\ea}{\end{eqnarray}}
\newcommand{\dif}{\mathrm{d}}

\usepackage{graphicx}
\usepackage{amssymb}
\usepackage{amsmath}
\usepackage[nosort]{cite}
\begin{document}
\thispagestyle{empty}
\begin{center}

\null \vskip-1truecm \vskip2truecm

{\Large{\bf \textsf{Angular Momentum in QGP Holography}}}

\vskip1truecm

{\large \textsf{Brett McInnes}}

\vskip0.1truecm
\textsf{\\ National
  University of Singapore}
  \vskip0.3truecm
\textsf{email: matmcinn@nus.edu.sg}\\

\end{center}
\vskip1truecm \centerline{\textsf{ABSTRACT}} \baselineskip=15pt
The quark chemical potential is one of the fundamental parameters describing the Quark-Gluon Plasma produced by sufficiently energetic heavy-ion collisions. It is not large at the extremely high temperatures probed by the LHC, but it plays a key role in discussions of the beam energy scan programmes at the RHIC and other facilities. On the other hand, collisions at such energies typically (that is, in peripheral collisions) give rise to very high values of the angular momentum density. Here we explain that holographic estimates of the quark chemical potential of a rotating sample of plasma can be very considerably improved by taking the angular momentum into account.

\medskip

\newpage
\addtocounter{section}{1}
\section* {\large{\textsf{1. Holography of the Quark Chemical Potential}}}
The quark matter phase diagram \cite{kn:ohnishi,kn:mohanty,kn:satz} will be explored in the beam energy scan programmes at various existing and projected facilities (RHIC, SHINE, FAIR, and NICA \cite{kn:ilya,kn:dong,kn:shine,kn:fair,kn:nica}). These scans involve collisions of heavy ions, producing \emph{Quark-Gluon Plasmas} (QGP) with large energy densities and correspondingly large values of the quark chemical potential, $\mu$, but relatively low values\footnote{The structure of the phase diagram is thought \cite{kn:katz,kn:enrodi} to be such that the QGP can exist at lower temperatures when $\mu$ is large than when it is small.} of the temperature, $T$. Studies of neutron stars could allow us to investigate even larger values of $\mu$, such that quark matter takes still more exotic forms, such as quark liquids \cite{kn:blaschke,kn:alford1,kn:frankfurt}. In both cases, an understanding of the values taken by the chemical potential is of basic importance in determining, for example, the equation of state of quark matter \cite{kn:eos}.

A theoretical understanding of the phase diagram is still under construction \cite{kn:race}. One important approach involves the use of \emph{gauge-gravity duality} \cite{kn:solana,kn:pedraza,kn:youngman,kn:gubser,kn:janik}, in which $\mu$ and $T$ are assumed to have interpretations in terms of the properties of a dual gravitational system: for example, of an electrically charged thermal asymptotically AdS black hole in the case of the QGP. In principle, this technique allows theoretical probes of the large-$\mu$ region of the diagram, supplementing other techniques, such as lattice methods \cite{kn:sharma}; for example, see \cite{kn:ewerz}. It also has important consequences for other aspects of holography; for example, the celebrated ``KSS bound'' on the viscosity/entropy density ratio for any fluid with an Einstein gravity dual \cite{kn:son,kn:schaefer} has to be modified at large values of $\mu$ \cite{kn:myers}.

In the holographic ``dictionary'', the chemical potential $\mu$ of the boundary field theory can be expressed in terms of the parameters of the dual charged black hole: specifically, in terms of the latter's electric charge and the radius $r_h$ of its event horizon. The temperature $T$ of the boundary field theory can likewise be expressed in terms of $r_h$ and the mass, and so can its energy density $\varepsilon$ (see\footnote{Specifically, the energy density $\varepsilon$ is dual to the mass of the black hole per unit horizon area; see Section 2.1 of \cite{kn:albash}, whose notation we adopt, for a detailed discussion of this point. Note that we are speaking here of a quantity measured infinitesimally (per unit area), not about the \emph{total} area of the event horizon, which of course has another interpretation in terms of entropy. In this work, again following \cite{kn:albash}, we shall always use such densities in order to avoid subtle global questions such as the topology of the event horizon or the large-scale structure of the plasma.} \cite{kn:ramallo}.) Thus, given $\mu$, $T$, and $\varepsilon$ for the fluid on the boundary, one has enough equations to fix the black hole parameters  $Q$, $r_h$, and $M$. (For example, see equations (\ref{DD}), (\ref{E}), and (\ref{M}) below; note however that the precise form of the equations does depend on the choice of the black hole.)

Now, physically, one does not expect $\mu$, $T$, and $\varepsilon$ to be completely uncorrelated: one often says, for example, that the region of the quark matter phase diagram to be explored in the beam energy scans is characterised by large values of the energy density \emph{and} the chemical potential. In fact, holography reflects this: $Q$, $r_h$, and $M$ are constrained by various interesting inequalities (see for example \cite{kn:dain,kn:hennig,kn:gwak}) and, of course, by the basic equation defining the position of the event horizon. Take, for example, a locally asymptotically AdS$_4$ electrically charged (AdS-Reissner-Nordstr$\ddot{\m{o}}$m) black hole with a spherical event horizon (leaving aside the possibility that the dual spatial geometry may not be realistic, a point we shall discuss in detail below). Then we have
\begin{equation}\label{A}
r_h^2\,\Big(1 + r_h^2\Big) - 2 Mr_h + {Q^2 \over 4\pi}\;=\;0,
\end{equation}
where we are temporarily setting $L$, the AdS curvature radius, equal to unity.

In view of our discussion above, a translation from the black hole parameters $Q$, $r_h$, and $M$ to the field theory parameters will convert this equation to a  simple holographic relation between $\mu$, $T$, and $\varepsilon$. The physical meaning of this relation is connected to the First Law of black hole thermodynamics: see \cite{kn:papasken} for a general survey of this, and \cite{kn:cognola}\cite{kn:gibperry}\cite{kn:klemm} for the precise details of the way it applies to the specific black holes we shall consider in this work\footnote{In fact, one uses conformity with the First Law to establish the physical values of the mass and the other parameters: see the discussion in Sections 2.1 and 2.2 of \cite{kn:gibperry}.}. It is then possible to solve for $\mu$ in terms of $T$ and $\varepsilon$, or in terms of the equivalent parameters $T$ and $U \equiv 2\pi [T + 2\varepsilon$]:
\begin{equation}\label{CHEMPOT}
4\pi \mu^2 \;=\; 1\;+\;{3 \over 2}\,\Big[U^2\;+\;U\sqrt{U^2 - 8}\;-\;4 \Big]\;-\;\pi T\,\Big[U\;+\sqrt{U^2 - 8}\Big].
\end{equation}

If we fix the temperature and energy density at some typical value for the QGP, as it is produced in (for example) the RHIC experiment \cite{kn:ilya}, then $\mu$ can be computed holographically. We find (see the end of the next section for the numerical details) that the values so obtained are \emph{not} realistic for these plasmas, even as to order of magnitude. (The estimated value exceeds 18000 MeV, whereas realistic values are below 1000 MeV.)

That is not in any way surprising, because the bulk geometry is excessively simple in this example. However, while it is true that current applications of gauge-gravity duality (see for example \cite{kn:hybrid}) are mainly based on using it as a qualitative guide, it would nevertheless be preferable to eliminate such extreme divergences between holographic computations and real data. One would at least like to know whether more complex versions of this toy model do indeed drive the predicted value of $\mu$ significantly in the downward direction.

The holographic approach can be made more realistic in two directions. One is based on adopting a more sophisticated gravity model in the bulk, and this is currently a subject of great interest \cite{kn:ficnar,kn:betz,kn:gyulassy}. The other, complementary, approach is to use more complex black hole spacetimes to model aspects of the QGP not previously considered. The most important of these aspects is the \emph{angular momentum} of the QGP, as it is produced in actual collisions. Taking this into account leads us to consider bulk black holes which themselves have angular momentum. Let us pursue this latter approach.

The angular momentum of an asymptotically AdS black hole does in fact have a holographic interpretation, arising from the well-known \emph{frame-dragging} effect (which, for AdS black holes, persists to infinity \cite{kn:cognola}). Frame-dragging at infinity, described by the usual parameter $a$ giving the angular momentum per unit mass of the bulk black hole, reproduces the motion of the fluid on the boundary ---$\,$ either as simple \emph{rotation} (first discussed in different applications in \cite{kn:sonner,kn:schalm}), or as a more complex \emph{shearing} \cite{kn:75,kn:76,kn:77,kn:shear}, with a non-trivial velocity profile of the kind studied in hydrodynamics\footnote{See \cite{kn:hydro} for the relevant hydrodynamic theory, \cite{kn:klemmhydro} for a holographic formulation.}; \emph{both} possibilities are realised by AdS black holes of various kinds. Thus we see that it is indeed possible to include angular momentum in the gauge-gravity description of the QGP.

Now actually all this is directly relevant to the observational situation, simply because the known or conjectured ways of producing quark matter \emph{do} normally involve very large angular momentum densities. The QGP produced in heavy-ion collisions will generically (that is, for peripheral collisions) acquire a very large angular momentum \cite{kn:liang,kn:bec,kn:huang} ---$\,$ as much as 7.2 $\times$ 10$^4$ (in natural units) per collision at the RHIC facility, and still more at the LHC. This angular momentum could, depending on details such as the viscosity of the fluid, take the form of either rotation or shearing \cite{kn:KelvinHelm,kn:viscous}; but, as we mentioned, one knows how to represent either possibility holographically. Similarly, quark matter in neutron stars will generically form the core of an object which rotates extremely rapidly \cite{kn:weber,kn:rotatingstar}. \emph{The possible role of angular momentum in holography must therefore be taken very seriously in applications}.

In this work, we shall focus primarily on the problem of constructing a holographic representation of a \emph{rotating} plasma with a given angular momentum/energy density ratio $a$. (See section 3 below for a brief discussion of the shearing plasma). The first step is to understand the correct way to introduce the parameter $a$ into the equations discussed earlier. For example, one generalizes equation (\ref{A}) by using a locally asymptotically AdS$_4$ electrically charged rotating (AdS-Kerr-Newman) black hole with a topologically spherical event horizon \cite{kn:carter,kn:hawrot}. If $M$, $Q$, $r_h$, and $a$ are as above, and $\Xi = 1 - a^2$, then we have\footnote{The factor involving $\Xi^2$ arises from demanding, as mentioned earlier, that the First Law should hold: see \cite{kn:cognola}\cite{kn:gibperry} for this particular case.}
\begin{equation}\label{B}
{r_h^2+a^2\over \Xi^2}\Big(1 + r_h^2 \Big) - 2 Mr_h + {Q^2 \over 4\pi}\;=\;0.
\end{equation}

Now if we again regard this as a holographic relation between $\mu$, $T$, and $\varepsilon$, we find that, formally, it is possible to regard $\mu$ as a function of $a$. That is, if we fix the temperature and energy densities at values characteristic of the QGP, then equation (\ref{B}) will force $\mu$ to vary as $a$ is varied. It turns out that, when viewed in this way, $\mu$ is an \emph{always decreasing}\footnote{Note that cosmic censorship ---$\,$ which in this case can be interpreted as the statement that the dual system is well-behaved thermodynamically, so that it has a well-defined entropy associated with an event horizon ---$\,$ does not require this behaviour, though of course it is consistent with it. Censorship demands that, if the angular momentum of the black hole increases, then (with other parameters fixed) its electric charge must \emph{eventually} decrease, so as to satisfy a certain inequality. The much stronger statement here is that $\mu$ always decreases as $a$ increases.} function of $a$. This opens up the appealing possibility that \emph{the inclusion of angular momentum can improve the holographic estimate} of $\mu$, even, perhaps, to the extent of rendering it fairly realistic.

In practice, the data are not known with sufficient precision for it to be fruitful to think of ``computing'' $\mu$. Instead, it is more meaningful to ask whether it is possible, given data on $T$ and $\varepsilon$, to find pairs ($\mu$, $a$) such that \emph{both} numbers are reasonably realistic under the conditions to be found in the beam scan experiments. This is of considerable theoretical interest, for otherwise one would run the risk of an inconsistency in applications of holography. For example, the modifications of the KSS bound proposed in \cite{kn:myers} involve values of $\mu/T$ which cannot be attained using reasonable parameter values in a simple (non-rotating) AdS-Reissner-Nordstr$\ddot{\m{o}}$m bulk geometry.

The question, then, is whether, using roughly realistic data, one can show that the effect of including angular momentum is large enough to achieve this goal. We shall see that, in the rotational case, it is; but that some additional effect is needed in the shearing case.

We begin by setting up a holographic model of a rotating plasma.

\addtocounter{section}{1}
\section* {\large{\textsf{2. Rotational Angular Momentum and The Chemical Potential}}}
Our objective is to set up a gauge-gravity model of heavy-ion collisions involving large \emph{rotational} angular momenta, and to use it to compute the chemical potential given data on the angular momentum and energy densities, together with the temperature.

The collision of two heavy nuclei is taken to occur along the $z$ axis of a Cartesian coordinate system. The internal motion of the plasma takes place in the $x-z$ planes, that is, the angular momentum generated in a peripheral collision is taken to lie along the $y$ axis. One assumes that the system can be studied by cutting the interaction zone into slices, with a thickness set by the nucleon length scale, perpendicular to the $y$ axis (that is, each $y$-section can be studied independently). This means that the system is effectively two-dimensional and is normally studied as such ---$\,$ see for example the diagrams in \cite{kn:liang,kn:bec,kn:huang,kn:KelvinHelm,kn:viscous}. A gauge-gravity model of such systems will therefore involve four-dimensional locally asymptotically AdS black holes\footnote{Of course, this means that energy and angular momentum densities will, in the first instance, take the form of quantities measured with respect to area rather than volume. This can be corrected by taking into account the thickness of the slices. That is, the actual three-dimensional densities of the plasma will be a universal constant multiple of the densities we use here. This will be taken into account in detail, below.}. The boundary is a three-dimensional conformally flat spacetime with spatial sections which can either be regarded as, or can approximate, the $x-z$ plane.

It is thought \cite{kn:KelvinHelm,kn:viscous} that the angular momentum transferred to the QGP in a peripheral collision can take the form either of \emph{rotation} or of internal \emph{shearing}; this depends on physical parameters such as the viscosity of the specific plasma being examined. In this section we shall construct a gauge-gravity model of the rotational case, returning briefly to the shearing case in the next section. Surprisingly, and very conveniently, it turns out that one can alternate between rotation and shearing simply by choosing the \emph{topology} of the event horizon.

As is well known, locally asymptotically AdS black holes can have various event horizon topologies \cite{kn:lemmo}. Consider the case of rotating, electrically charged AdS black holes with topologically \emph{spherical} event horizons. Here the angular velocity is constant (both in space and in time) on the event horizon but also on the boundary, so there is no differential motion there: the boundary is \emph{rotating} \cite{kn:sonner}\cite{kn:schalm}, not shearing. This is the kind of black hole we need in this section. (The rotational motion of quark plasmas has been studied in, for example, \cite{kn:rotation}\cite{kn:csorgo}, though not from the holographic point of view.)

In attempting to construct a gauge-gravity model of this situation, one immediately encounters an obvious difficulty: the QGP exists in a space which is flat, not spherical like the spatial sections of the black hole boundary. However, a sufficiently large sphere, or deformed sphere, can be used to approximate the finite domain in which the plasma exists. If we want to use the topologically spherical black hole to model the rotating plasma, then we need to reassure ourselves that the deformed sphere defined at infinity by the rotating black hole is indeed sufficiently large.

The four-dimensional topologically spherical AdS-Kerr-Newman metric takes the form discovered by Carter \cite{kn:carter,kn:hawrot},
\begin{equation}\label{C}
g(\m{AdSKN}) = - {\Delta_r \over \rho^2}\Bigg[\,\m{d}t \; - \; {a \over \Xi}\m{sin}^2\theta \,\m{d}\phi\Bigg]^2\;+\;{\rho^2 \over \Delta_r}\m{d}r^2\;+\;{\rho^2 \over \Delta_{\theta}}\m{d}\theta^2 \;+\;{\m{sin}^2\theta \,\Delta_{\theta} \over \rho^2}\Bigg[a\,\m{d}t \; - \;{r^2\,+\,a^2 \over \Xi}\,\m{d}\phi\Bigg]^2,
\end{equation}
where
\begin{eqnarray}\label{D}
\rho^2& = & r^2\;+\;a^2\m{cos}^2\theta \nonumber\\
\Delta_r & = & (r^2+a^2)\Big(1 + {r^2\over L^2}\Big) - 2\Xi^2 Mr + {\Xi^2Q^2 \over 4\pi}\nonumber\\
\Delta_{\theta}& = & 1 - {a^2\over L^2} \, \m{cos}^2\theta \nonumber\\
\Xi & = & 1 - {a^2\over L^2},
\end{eqnarray}
and where we reinstate the asymptotic curvature radius $L$ for later convenience. Here $a$ (which has units of length) can be taken to be positive; and we notice that we must have $a\,<\,L$. This bound on the angular momentum in terms of the asymptotic curvature scale is imposed by the geometric requirement that $\Delta_{\theta}$ should have a consistent sign (and \emph{not} by censorship). Intuitively, one expects that, in the application to the very rapidly rotating plasmas produced in peripheral heavy-ion collisions, $a/L$ will be close to the maximal value; and so it will prove.

The AdS Kerr-Newman coordinate $t$ does not represent proper time in the bulk. But the boundary geometry can be represented by a metric of the form
\begin{equation}\label{EGAD}
g({AdSKN})_{\infty}\;=\;-\,\m{d}t^2 \;-\;{2a\,\m{sin}^2(\theta)\,\m{d}t \m{d}\phi\over \Xi} \;+\; {L^2 \, \m{d}\theta^2 \over 1 - (a/L)^2\m{cos}^2(\theta)} \;+\; {L^2 \m{sin}^2(\theta)\m{d}\phi^2\over \Xi},
\end{equation}
from which we see that, if we fix the conformal gauge in this manner, $t$ can be taken to represent proper time at infinity. However, if we do this, then the spatial geometry at infinity is not precisely spherical: it is that of a \emph{deformed} sphere. Nevertheless, the geometry is spherical, if we are sufficiently close to one of the poles; but the radius of that sphere is \emph{not} equal to $L$.

To see how this works, take a circle centred on the pole $\theta = 0$, and compute the ratio of its circumference ($2\pi L\, \m{sin}(\theta)/\sqrt{\Xi}$) to its radius ($\int_0^{\theta}L\,\m{d}\theta /\sqrt{1-(a/L)^2\m{cos}^2(\theta)}$); one finds that, as the radius tends to zero, this ratio tends to 2$\pi$, that is, the space is regular (and approximately spherical), only because of the presence of $\Xi$ in the final term of equation (\ref{EGAD}). This is why that factor \emph{must} be present. But it then follows that, even near the pole, the spatial geometry is approximately that of a round two-sphere of radius $\hat{L} = L/\sqrt{\Xi}$.

Our problem is that the actual space in which the QGP exists is flat, not spherical; and the spatial geometry in the core of a neutron star, while not exactly flat, is also not spherical. We need $\hat{L}$ to be sufficiently large for it to be possible to approximate the sphere we have been discussing by its tangent plane at the pole. To see whether this is possible, we proceed as follows.

Consider a massive particle at infinity in this geometry, with zero angular momentum. Its worldline has unit tangent $\dot{t}\partial_t + \dot{\phi}\partial_{\phi}$, where the dot denotes differentiation with respect to proper time. The inner product of this tangent vector with the Killing vector $\partial_{\phi}$ vanishes, and this yields $\dot{\phi}$ = $a\,\dot{t}/L^2$, showing that the particle is frame-dragged in the $\phi$ direction at an angular velocity of $\omega$ = $a/L^2$. Therefore, $\hat{L}$ can be computed in the holographic picture if one has data on $a$ and can determine the angular velocity $\omega$ of the plasma. In principle, we could compute $\omega$ from $a$ given sufficient information on the structure of a given specimen of rotating plasma. In practice, phenomenological models of the rotating plasma allow one to estimate the (dimensionless) linear velocity of a plasma sample at a known radial distance from the centre of rotation, and the angular velocity can be computed straightforwardly, allowing for relativistic effects, from these quantities.

If $\hat{L}$ is large compared to the radius of the rotating plasma specimen, $\varrho_{max}$, then the sphere can indeed be used to approximate the planar geometry. In that case, defining $\varrho$ = $\hat{L}\theta$, one can interpret the coordinates $\varrho$ and $\phi$ as plane polar coordinates in the $x-z$ plane described earlier. The condition for these approximations to be good is that sin($\theta$) should be well approximated by $\theta$, that is, that $(\varrho_{max}/\hat{L})^3$ should be negligible.

In order to be specific, we use as our standard example the model discussed in \cite{kn:KelvinHelm}. In the rotating case, one finds there that
the maximal dimensionless speed (at the outer edge of the plasma) can be roughly estimated at a value of the order of 0.25, and the corresponding radius is around 6 femtometres; also, (see below) we will use an estimate for $a$ of around 20 fm. All of these numbers are given so as to provide intuition: we claim only that they are reasonable, not that they are fully realistic\footnote{It might be objected that some alternative choices will lead to values such that $a/L > 1$. That is correct, but the conclusion in that case would be that holographic methods simply do not apply. In our view, the uncertainty in the data is such that this is not a matter of concern at present.}.

One then obtains $L$ $\approx$ 21 fm ---$\,$ around a typical nuclear physics scale, and a little larger than $a$; so with these data, the $a/L < 1$ bound is respected, but not by a large margin. What we conclude from this is not that we need to use these specific data, but rather that $a/L$ is typically close to unity under these circumstances. That will suffice for our purposes.

With these choices, $\hat{L}$ is indeed significantly larger than the radius of the plasma sample; it is around 53 fm, and $(\varrho_{max}/\hat{L})^3$ is no larger than about 0.0015. Thus we have here a way of constructing a holographic representation of a plasma rotating in an approximately \emph{flat} space.

The quark-matter cores of neutron stars, if they exist, consist of some other form of quark matter than a plasma; but it is interesting to see what happens in that case if we ignore this. One has, for a very high-frequency pulsar, a rotational frequency around 700 Hz \cite{kn:pulsar}; the core radius is of order 1 km. Following the same procedure as above, one finds that $\hat{L}$ is actually smaller than the radius of the core; which means that the spherical geometry cannot be used to approximate (say) the equatorial section of the neutron star. Admittedly, in the intense gravitational field of the core, the geometry is not exactly flat, but nor does it resemble the geometry of (a large portion of) a deformed two-sphere. We conclude that rotating AdS black holes could not be used to describe the effects of large angular momenta on the quark matter in the cores of rapidly rotating neutron stars, even in the unlikely event that the latter should prove to be composed of some kind of quark matter similar to the QGP. Henceforth we confine our attention to the application to peripheral heavy-ion collisions.

Now we can relate the boundary parameters to those of the black hole. The area of the event horizon is $4\pi (r_h^2 \,+\,a^2)/\Xi$, so that the energy density of the black hole at its event horizon is
\begin{equation}\label{DD}
\varepsilon \;=\;{[1-(a^2/L^2)] M \over 4\pi (r_h^2 \,+\,a^2)}.
\end{equation}

The Hawking temperature in this case \cite{kn:cognola} is
\begin{equation}\label{E}
T\;=\;{r_h \Big(1\,+\,a^2/L^2\,+\,3r_h^2/L^2\,-\,(a^2\,+\,Q^2/4\pi)/r_h^2\Big)\over 4\pi (a^2\,+\,r_h^2)}.
\end{equation}
The quantities $\varepsilon$ and $T$ are interpreted holographically in terms of the energy density and temperature of the fluid we are studying at infinity. All that remains is to find the relationship between the field theory chemical potential and the bulk charge parameter. This is most easily derived in the Euclidean version of the geometry.

The topologically spherical Euclidean AdS-Kerr-Newman gravitational instanton metric takes the form
\begin{equation}\label{F}
g(\m{EAdSKN}) = {\Delta^E_r \over \rho_E^2}\Bigg[\,\m{d}t \; + \; {a \over \Xi^E}\m{sin}^2\theta \,\m{d}\phi\Bigg]^2\;+\;{\rho_E^2 \over \Delta^E_r}\m{d}r^2\;+\;{\rho_E^2 \over \Delta^E_{\theta}}\m{d}\theta^2 \;+\;{\m{sin}^2\theta \,\Delta^E_{\theta} \over \rho_E^2}\Bigg[a\,\m{d}t \; - \;{r^2\,-\,a^2 \over \Xi^E}\,\m{d}\phi\Bigg]^2,
\end{equation}
where
\begin{eqnarray}\label{eq:G}
\rho_{E}^2& = & r^2\;-\;a^2\m{cos}^2\theta \nonumber\\
\Delta^E_r & = & (r^2-a^2)\Big(1 + {r^2\over L^2}\Big) - 2(\Xi^E)^2 Mr - {(\Xi^E)^2Q^2 \over 4\pi}\nonumber\\
\Delta^E_{\theta}& = & 1 + {a^2\over L^2} \, \m{cos}^2\theta \nonumber\\
\Xi^E & = & 1 + {a^2\over L^2}.
\end{eqnarray}
Here the superscript or subscript E denotes the Euclidean version of the respective quantity.
When cosmic censorship is satisfied, the polynomial $\Delta^E_r$ has a positive root r$_0$, and the range of $r$ is constrained by $r \geq r_0$; note that, from the definition of $\Delta^E_r$, $r^2_0 > a^2$, so $\rho_E^2$ is always positive.

The Euclidean electromagnetic potential is given by
\begin{equation}\label{H}
A^E \;=\;\Bigg[-{\Xi^E Qr \over 4\pi \rho_E^2}\;+\;\kappa_t \Bigg]\m{d}t\;-\;\Bigg[{aQ r\,\m{sin}^2\theta \over 4\pi \rho_E^2}\;+\;\kappa_{\phi}\Bigg]\m{d}\phi;
\end{equation}
here $\kappa_t$ and $\kappa_{\phi}$ are constants. Notice that, through $\rho_{E}$, the coefficients here depend on both $r$ and $\theta$.

The Euclidean squared length of the Killing vector $\partial_t$ is given by
\begin{equation}\label{I}
g(\m{EAdSKN})(\partial_t,\partial_t) \;=\; {\Delta^E_r \;+\;a^2\,\m{sin}^2\theta \,\Delta^E_{\theta} \over \rho_E^2}.
\end{equation}
The geometry being Euclidean, this means that $\partial_t$ itself must vanish when $\Delta_r^E$ vanishes and $\theta$ is either zero or $\pi$, and hence $A^E$ must satisfy $A^E_t$ = $A^E(\partial_t)$ = 0 at those locations (which correspond to the points where the ergosurface touches the event horizon in the Lorentzian version of the geometry). Similarly $\partial_{\phi}$ vanishes at the poles and so $A^E_{\phi}$ has to vanish there.

Substituting $r =  r_0$ and $\theta$ = 0 or $\pi$ into the potential, it follows that $\kappa_t$ is given by $\kappa_t \;=\;\Xi^E Qr_0/4\pi (r_0^2 - a^2)$ and $\kappa_{\phi} = 0$, so that we have
\begin{equation}\label{J}
A^E \;=\;\Bigg[-{\Xi^E Qr \over 4\pi \rho_E^2}\;+\;{\Xi^E Qr_0 \over 4\pi (r_0^2 - a^2)}\Bigg]\m{d}t\;-\;{aQ r\,\m{sin}^2\theta \over 4\pi \rho_E^2}\m{d}\phi.
\end{equation}
Returning to the Lorentzian signature, we therefore have an electromagnetic potential at infinity given by
\begin{equation}\label{K}
A_{\infty} \;=\;{\Xi Qr_h \over 4\pi (r_h^2 + a^2)}\m{d}t.
\end{equation}

We see from equation (\ref{EGAD}) that $\partial_t$ is a unit vector at infinity, and so it is the tangent vector to the worldline of a (stationary) observer there. This observer therefore sees an electric potential given simply by $A_{\infty}(\partial_t)$. This dimensionless quantity gives us \cite{kn:koba} the dimensionless version of the field theory chemical potential, $\mu L$, so we have
\begin{equation}\label{M}
\mu \;=\;{[1-(a^2/L^2)] Qr_h \over 4\pi L [r_h^2 + a^2]}.
\end{equation}
The form of the numerator in this expression suggests that high values of $a$ (relative to $L$) will suppress the value of $\mu$. On the other hand, however, we have, from equations (\ref{DD}) and (\ref{M}),
\begin{equation}\label{N}
{\mu\over \varepsilon}\;=\;{r_h\over L}\times{Q\over M}.
\end{equation}
We see that $\mu/\varepsilon$, a quantity which refers to the properties of the boundary fluid, is proportional to the corresponding black hole charge-to-mass ratio. We see also, however, that $\mu/\varepsilon$ need not be small even if $Q/M$ were so; one needs also to determine whether the black hole is ``large'' (in the usual sense, that is, $r_h/L$ is large). Thus it is not clear that $\mu$ must be small relative to the other parameters, and indeed it often is not. This is our problem.

As discussed in the preceding section, all of the parameters are related by the equation which locates the event horizon,
\begin{equation}\label{BB}
{r_h^2+a^2\over \Xi^2}\Bigg(1 + {r_h^2\over L^2}\Bigg) - 2 Mr_h + {Q^2 \over 4\pi}\;=\;0.
\end{equation}
Substituting equation (\ref{DD}) into equation (\ref{BB}), we have
\begin{equation}\label{O}
{r_h^2+a^2\over \Xi}\Bigg({1 + {r_h^2\over L^2}\over \Xi} - 8\pi \varepsilon r_h \Bigg)  + {Q^2 \over 4\pi }\;=\;0.
\end{equation}
Combining this with equation (\ref{E}), we have a pair of equations which, for given values of $T$, $\varepsilon$, $L$, and $a$, can be solved to find $Q$ and $r_h$. One can then compute $M$, if desired, from equation (\ref{DD}); more importantly, $\mu$ can now be computed from equation (\ref{M}). We can fix $T$, $L$, and $\varepsilon$ at values typical of conditions in heavy-ion collisions, and then explore how $\mu$ varies with $a$.

It will be convenient to express all of the parameters in dimensionless form (denoted by a tilde), by multiplying or dividing by $L$: the two basic equations (\ref{E}) and (\ref{O}) then become
\begin{equation}\label{P}
\tilde{T}\;=\;{\tilde{r}_h \Big(1\,+\,\tilde{a}^2\,+\,3\tilde{r}_h^2\,-\,(\tilde{a}^2\,+\,\tilde{Q}^2/4\pi)/\tilde{r}_h^2\Big)\over 4\pi (\tilde{a}^2\,+\,\tilde{r}_h^2)}
\end{equation}
and
\begin{equation}\label{Q}
{\tilde{r}_h^2+\tilde{a}^2\over \Xi}\Bigg({1 + \tilde{r}_h^2\over \Xi} - 8\pi \tilde{\varepsilon} \tilde{r}_h \Bigg)  + {\tilde{Q}^2 \over 4\pi }\;=\;0,
\end{equation}
where of course now $\Xi = 1 - \tilde{a}^2$.

We can now proceed, using any convenient system of units; in the application to heavy ion collisions, the natural units are femtometres or MeV. A typical energy density for the plasma produced in a heavy ion collision is roughly \cite{kn:phobos} 3000 MeV/fm$^3$, or about 15 fm$^{-4}$. The \emph{maximum} angular momentum density in the RHIC experiments has been estimated \cite{kn:75} at around 360 fm$^{-3}$, which leads to an estimate of $a_{max}$ $\approx$ 24 fm, so we can assume that typical collisions will have values of $a$ around 20 fm. In practice, since the holographic model requires $\tilde{a} < 1$, and since our earlier discussion suggests that $\tilde{a}$ is in fact just below unity, we proceed by examining a range of such values for $\tilde{a}$ . The thickness of the sections here is of the order of 2 fm, giving us rough estimates for $\tilde{\varepsilon}$. A typical temperature for the QGP (say, near to the current estimated position of the quark matter critical point) is around 200 MeV, or roughly 1 fm$^{-1}$, and we use this to estimate $\tilde{T}$. As mentioned earlier, the beam energy scans may generate plasmas, at high values of $\mu$, with somewhat lower temperatures, so we also consider a lower value $T$ $\approx$ 100 MeV. We stress again that precision is not to be looked for here: we claim only that none of these numbers is unreasonable.

With these data, we can (numerically) solve\footnote{For the data we use, these equations always prove to have a (unique) pair of real solutions. It should be noted however that there must exist data that will lead to a pair of equations with no real solutions; we know this because it is always possible, formally, to violate cosmic censorship. In other words, the holographic approach does not work for completely arbitrary data.} equations (\ref{P}) and (\ref{Q}) for $\tilde{r}_h$ and $\tilde{Q}$; substituting the results into equation (\ref{M}) we obtain the following results (expressed in terms of the usual units, MeV) for the chemical potential.

When $\tilde{a}$ $\approx$ 0, one finds that $\mu$ is predicted to be well over 18000 MeV, for either choice of temperature. To put this in perspective, a typical guess for the location of the quark matter critical point would put it at around\footnote{The nuclear physics literature normally uses the baryonic chemical potential; here $\mu$ is the quark chemical potential, so we have to correct for this.} $\mu \approx$ 150 MeV; the beam scans are expected to explore a range of values up to a few times this.

When $\tilde{a}$ is chosen to lie between 0.90 and 0.98, however, the same calculation yields very different results: see the table.
\begin{center}
\begin{tabular}{|c|c|c|c|c|c|}
  \hline
 & $\tilde{a}$ = 0.90  &  $\tilde{a}$ = 0.92 & $\tilde{a}$ = 0.94 &$\tilde{a}$ = 0.96 &$\tilde{a}$ = 0.98 \\
\hline
$T$ = 200 MeV & 2570  & 1732 & 1017 &  466 & 116 \\
$T$ = 100 MeV &  2588 & 1747 & 1029 & 474 &  120 \\
\hline
\end{tabular}
\end{center}
One sees, first, that there is very little variation with temperature, so the lower temperatures associated with the beam scans will not affect the situation materially; secondly, that angular momentum is very effective in improving the alignment of theory with data, reducing the predicted value of $\mu$ by large factors. Values of $\tilde{a}$ in the range 0.94 - 0.98 are needed, but these are by no means unrealistic.

It was by no means obvious that we would reach this conclusion. Proof of this is provided in the next section, where we find that, in a model of the \emph{shearing} QGP, $\mu$ varies with the angular momentum so slowly that no physical value of the latter can reduce the former to a reasonable range of values.

Perhaps the best way to state the case is to say that including angular momentum is a \emph{necessary} component of a quasi-realistic theoretical description of the QGP at relatively large values of $\mu$; in particular, that, when using holography to describe such plasmas, one should perhaps consider black holes with non-zero angular momentum as the default choice of bulk geometry.

For example, Myers et al. \cite{kn:myers} show that, for a black hole described by a four-derivative action in the bulk, the KSS computation of the viscosity-to-entropy density has to be corrected in the presence of a chemical potential, yielding
\begin{equation}\label{MYERS}
{\eta\over s}\;=\;{1\over 4\pi}\,\Bigg[1\;-\;8c_1 \;+\;{16\bar{\mu}^2\over 3\Big(1\;+\;\sqrt{1\;+\;2\bar{\mu}^2/3}\Big)^2}\,\times\,\Big(c_1\;+\;6c_2\Big)\Bigg],
\end{equation}
where $\bar{\mu} = \mu/T$, $c_1$ is the coupling for the contribution to the gravitational action of the form $R_{abcd}R^{abcd}$, and $c_2$ is the coupling for $R_{abcd}F^{ab}F^{cd}$, where $F_{ab}$ is the field strength tensor. The parameters $c_1$ and $c_2$ are very small; but the dependence on $\bar{\mu}$ means that discernible deviations from the KSS bound (both upwards and downwards, depending on the signs of $c_1$ and $c_2$) are in principle possible for sufficiently large chemical potentials. (The need to take such higher-order terms into account in applied holography has recently been emphasised in \cite{kn:ficnar,kn:betz}.)

Equation (\ref{MYERS}) is obtained by using a black hole background with zero angular momentum. Our results suggest that this whole question has to be re-considered: under these circumstances, one should be using black holes endowed with
substantial angular momenta, like the AdS-Kerr-Newman black hole or a suitable generalization of it, to compute ${\eta /s}$. Otherwise one would be using values of $\mu$ which might not be consistent with the model itself.

We should bear in mind, however, that this entire discussion pertains to the situation in which the angular momentum of the plasma is associated with overall \emph{rotation}. We now turn to the equally important, if mathematically less familiar, case of a plasma which has a large internal angular momentum due to its \emph{shearing} motion \cite{kn:liang,kn:bec,kn:huang,kn:KelvinHelm,kn:viscous}.

\addtocounter{section}{1}
\section* {\large{\textsf{3. Shearing Angular Momentum and The Chemical Potential}}}
To describe a shearing fluid at infinity, we cannot use topologically spherical black holes. Fortunately there is another, entirely different class of locally asymptotically AdS black holes, with \emph{planar} event horizon topologies \cite{kn:lemmo}, and it turns out that these are precisely what we need.

The four-dimensional Planar AdS-Reissner-Nordstr$\ddot{\m{o}}$m black hole metric is given \cite{kn:77} by
\begin{equation}\label{ALPHA}
g(\m{PAdSRN}) = -\, \Bigg[{r^2\over L^2}\;-\;{8\pi M^*\over r}+{4\pi Q^{*2}\over r^2}\Bigg]dt^2\; + \;{dr^2\over {r^2\over L^2}\;-\;{8\pi M^*\over r}+{4\pi Q^{*2}\over r^2}} \;+\;r^2\Big[d\psi^2\;+\;d\zeta^2\Big],
\end{equation}
where $\psi$ and $\zeta$ are (dimensionless) coordinates on the plane, where $L$ is the asymptotic AdS curvature radius, and where $M^*$ and $Q^*$ are parameters which allow us to compute the mass and charge densities on the horizon: the densities are given by $M^*/r_h^2$ and $Q^*/r_h^2$, where as usual $r = r_h$ locates the event horizon.

One can use this geometry to give a holographic estimate of the quark chemical potential in the absence of angular momentum. As one would expect, the result is similar to the value we obtained in the preceding section, about two orders of magnitude larger than the physical values; so we have the same problem as before.

Let us now add angular momentum, measured as usual by the parameter $a$: it is now the ratio of the angular momentum density on the event horizon to its mass density. It turns out that the corresponding black holes differ from their topologically spherical counterparts in one crucial particular: the angular velocity is still constant on the event horizon, but \emph{not at infinity}. Instead, the metric at infinity takes the ``Peripheral Collision'' form \cite{kn:75}:
\begin{equation}\label{DELTA}
g_{\rm PC} \;=\; - \, \m{d}t^2 \;-\; 2\omega_{\infty}(x)\, L \, \m{d}t\m{d}z \;+\; \m{d}x^2 \;+\; \m{d}z^2.
\end{equation}
Here $x$ and $z$ are Cartesian coordinates, related in a simple way to the coordinates $\psi$ and $\zeta$ in equation (\ref{ALPHA}). The function $\omega_{\infty}(x)$ is the asymptotic value of the angular velocity of the black hole, and it is in general a non-trivial function of $x$. Free particles, with $x$ = constant and zero momentum, are frame-dragged in the $z$ direction at a dimensionless speed given by
\begin{equation}\label{EPSILON}
v(x) \; \equiv \; \m{d}z/\m{d}t = \omega_{\infty}(x)L.
\end{equation}
This function describes the shearing motion within the plasma. In principle \cite{kn:chrusc1,kn:chrusc2,kn:delay} it may be possible to prescribe it arbitrarily; in practice, only a few locally asymptotically AdS solutions of the Einstein equations are actually known explicitly in this case, and we only consider those. (In fact, these solutions are sufficient, in the sense that they represent the broad possibilities for the general shape of the velocity profile within a shearing plasma, and the actual profile is currently not known more accurately in any case.)

\subsubsection*{{\textsf{3.1 Shearing Near the Axis}}}
The first of these solutions (to the Einstein equations with a negative cosmological constant) is the metric we have called the ``KMV$_0$ metric'', obtained in the zero-charge case by Klemm, Moretti, and Vanzo \cite{kn:klemm}; with the addition of electric charge, we call these the ``QKMV$_0$ metrics'':
\begin{equation}\label{BETA}
g(\m{QKMV_0}) = - {\Delta_r\Delta_{\psi}\rho^2\over \Sigma^2}\,dt^2\;+\;{\rho^2 \over \Delta_r}dr^2\;+\;{\rho^2 \over \Delta_{\psi}}d\psi^2 \;+\;{\Sigma^2 \over \rho^2}\Bigg[\omega\,dt \; - \;d\zeta\Bigg]^2,
\end{equation}
where the coordinates and $L$ are as in equation (\ref{ALPHA}), and where
\begin{eqnarray}\label{GAMMA}
\rho^2& = & r^2\;+\;a^2\psi^2 \nonumber\\
\Delta_r & = & a^2+ {r^4\over L^2} - 8\pi M^* r + 4\pi Q^{*2}\nonumber\\
\Delta_{\psi}& = & 1 +{a^2 \psi^4\over L^2}\nonumber\\
\Sigma^2 & = & r^4\Delta_{\psi} - a^2\psi^4\Delta_r\nonumber\\
\omega & = & {\Delta_r\psi^2\,+\,r^2\Delta_{\psi}\over \Sigma^2}\,a.
\end{eqnarray}

Here (and in the metric used in Section 3.2 below) the parameters $M^*$ and $Q^*$ correspond to the \emph{physical} mass and charge (densities), in the same sense as discussed by Gibbons et al. \cite{kn:gibperry}: that is, they are the parameters which respect the First Law of black hole thermodynamics. This was confirmed by Klemm et al. in Section IV.B of \cite{kn:klemm}.

Clearly equation (\ref{BETA}) reduces to (\ref{ALPHA}) when $a$ = 0, so this is the generalization of the Planar AdS-Reissner-Nordstr$\ddot{\m{o}}$m geometry to allow for the presence of angular momentum. (The space with $r$ = constant, $t$ = constant, described by coordinates ($\psi, \zeta$) still has planar \emph{topology}, though it is no longer flat except at infinity.)

The velocity profile at infinity in this specific case takes the form
\begin{equation}\label{ZETA}
v(x) \;=\; a\psi^2/L.
\end{equation}
(Notice that this equation implies that $a > 0$, since $v(x)$ is taken to be positive away from the axis.) When expressed in terms of $x$ instead of $\psi$, this has the form of the square of the ``lemniscatic sine'' function \cite{kn:76}; for certain parameter values, it has the shape shown in Figure 1. This is a typical fluid shearing profile for the part of the fluid which is near to the axis along which the velocity vanishes.

\begin{figure}[!h]
\centering
\includegraphics[width=0.55\textwidth]{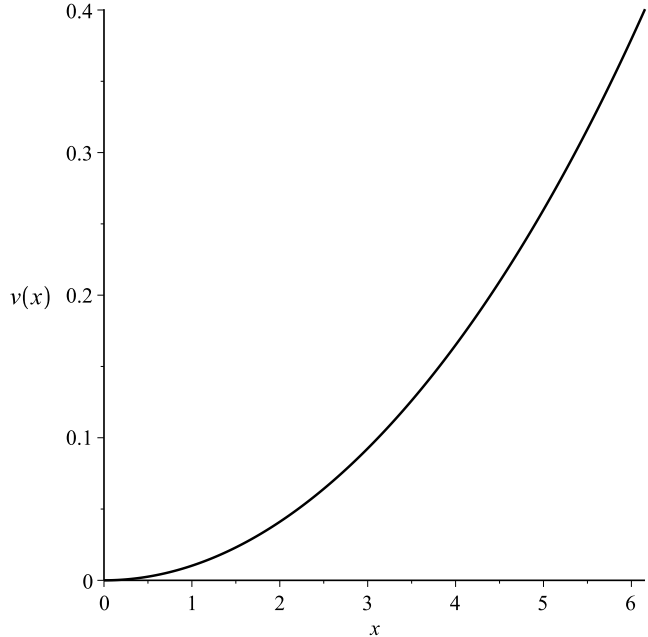}
\caption{QKMV$_0$ velocity profile, $a > 0$.}
\end{figure}

Finally, the electromagnetic potential form associated with such a black hole takes the form
\begin{equation}\label{ETA}
A = -{Q^*r \over  \rho^2}\m{d} t \;-\; {aQ^*r\psi^2 \over  \rho^2}\m{d}\zeta.
\end{equation}

Of course, the behaviour shown in Figure 1 cannot persist for larger values of $x$, if for no other reason than causality; the curve must eventually bend over, so that the velocity is bounded. There is in fact another black hole geometry with a velocity profile describing the motion at greater distances from the axis; we will discuss it later.

As in the preceding section, we now need a set of equations relating the energy density, temperature, and chemical potential to the corresponding quantities in the QKMV$_0$ geometry.

First, note that since the metric induced on $t$ = constant sections at the horizon still has determinant equal to $r^4_h$, it follows that $M^*$ and $Q^*$ retain their interpretations as in (\ref{ALPHA}), so that, in particular, the energy density at the horizon is still given by
\begin{equation}\label{VARDELTA}
\varepsilon \;=\;M^*/r^2_h.
\end{equation}
The Hawking temperature can be expressed in terms of  $M^*$ and $r_h$ only (as in \cite{kn:klemm}):
\begin{equation}\label{VARVARDELTA}
T\;=\;{r_h^3/L^2\;-\; 2\pi M^* \over \pi r_h^2}.
\end{equation}

Using equation (\ref{ETA}), we can compute \cite{kn:77} the chemical potential of the field theory in much the same manner as in the preceding section, obtaining finally
\begin{equation}\label{VARVARVARDELTA}
\mu \;=\;{Q^*r_h \over L [r_h^2 + aL]}.
\end{equation}
This is quite different from the corresponding formula in the rotational case (equation (\ref{M})): in particular, it involves $a$ itself (which, it will be recalled, is positive here) and not just its square; more crucially, the factor involving $\Xi$ is absent, so it is less clear that large angular momenta suppress the chemical potential here. The formula analogous to equation (\ref{N}) is
\begin{equation}\label{OMICRON}
{\mu\over \varepsilon}\;=\;{r_h\over L}\times{Q^*\over M^*}\times {1 \over 1\,+\,[aL/r_h^2]}.
\end{equation}

In \cite{kn:shear}, we gave a detailed discussion of the relevant parameter values, based on observational data from the RHIC experiment. We will use those values here and in the next section. The dimensionless parameters we need are computed by using appropriate multiples and quotients of $L$. (The latter is necessarily determined in an entirely different way from the rotating case, and differs from the value used earlier: see \cite{kn:shear} for the details; here $L \approx 11$ fm.) The results are as follows: $\tilde{\varepsilon} \approx 300,\, \tilde{T} \approx 11,\, \tilde{a} \approx 1.36$. (Note that values of $\tilde{a}$ above unity are no longer forbidden here.)

One quickly finds that the situation in this case differs quite drastically from our results in the preceding section. Combining equations (\ref{VARDELTA}) and (\ref{VARVARDELTA}), we obtain
\begin{equation}\label{PI}
r_h\;=\;\pi L^2\,[T\;+\;2\varepsilon],
\end{equation}
and from this one sees that the dimensionless version of $r_h$, $\tilde{r}_h$, can be computed using only $\tilde{T}$ and $\tilde{\varepsilon}$; remarkably, once those parameters are fixed, it does not depend on the angular momentum. One finds that it is actually quite large: $\tilde{r}_h \approx 1920$.

The dimensionless version of equation (\ref{VARVARVARDELTA}) takes the form
\begin{equation}\label{RHO}
\tilde{\mu} \;=\;{\tilde{Q}^*\tilde{r}_h \over  [\tilde{r}_h^2 + \tilde{a}]}.
\end{equation}
Since $\tilde{r}_h^2$ is so large compared to $\tilde{a} \approx 1.36$, one begins to suspect that $\mu$ is not much affected by variations in the amount of angular momentum.

To confirm that, we turn to the definition of $\tilde{r}_h$: it is the largest real solution of the equation
\begin{equation}\label{SIGMA}
\tilde{a}^2\;+\;4\pi \tilde{Q}^{*2} \; - \; 8\pi \tilde{M}^*\tilde{r}_{h} \; + \;\tilde{r}_{h}^4 \; = \; 0.
\end{equation}
Eliminating $\tilde{Q}^*$ and $\tilde{M}^*$, one can write this as
\begin{equation}\label{VARSIGMA}
\tilde{a}^2\;+\;{4\pi \tilde{\mu}^2 \,[\tilde{r}^2_h + \tilde{a}]^2 \over \tilde{r}^2_h}\; - \; 8\pi \tilde{\varepsilon} \tilde{r}^3_{h} \; + \;\tilde{r}_{h}^4 \; = \; 0,
\end{equation}
or
\begin{equation}\label{VARVARSIGMA}
\tilde{\mu}\;=\;\sqrt{{(8\pi \tilde{\varepsilon} \tilde{r}^3_{h} \; - \;\tilde{r}_{h}^4 \; -\;\tilde{a}^2)\,\tilde{r}_h^2\over 4\pi (\tilde{r}^2_h + \tilde{a})^2}}.
\end{equation}
Since $\tilde{r}_h$ is determined only by $\tilde{T}$ and $\tilde{\varepsilon}$, this equation gives the explicit dependence of $\tilde{\mu}$ on $\tilde{a}$ when the temperature and energy density are fixed. We see at once that, as one would hope, it is a decreasing function. However, for physical values of the parameters, it decreases extremely slowly, and is effectively constant for reasonable values of the angular momentum. That constant translates to around 84.2 fm$^{-1}$ or about 16600 MeV, which is still unphysical.

Thus we see that the ability of angular momentum to solve the problem in the rotational case was not foreordained: it works in that case, but not here.

\subsubsection*{{\textsf{3.2 Shearing Far From the Axis}}}
Now we turn to the other family of explicitly known asymptotically AdS charged planar black holes which induce a shearing motion at infinity. These differ from the QKMV$_0$ metrics by depending on a new positive parameter $\ell$ (with units of length), which is in some ways analogous to NUT charge. The ``$\ell$QKMV$_0$ metrics'' were introduced in \cite{kn:shear} (as members of the very general Pleba\'nski--Demia\'nski family of metrics \cite{kn:plebdem,kn:grifpod}), to which we refer the reader for the details.

The metrics take the form
\be
\label{TAU}
g(\ell {\rm QKMV}_0)=-\frac{\Delta_r\Delta_\psi\rho^2}{\Sigma^2}\,\dif t^2+\frac{\rho^2\dif r^2}{\Delta_r}+\frac{\rho^2\dif\psi^2}{\Delta_\psi}+\frac{\Sigma^2}{\rho^2}\left[\omega\dif t-\dif\zeta\right]^2,
\ee
where
\ba
\label{UPSILON}
\rho^2&=&r^2+(\ell+a\psi)^2\cr
\Delta_r&=&\frac{(r^2+\ell^2)^2}{L^2}-8\pi M^*r+a^2+4\pi Q^{*2}\cr
\Delta_\psi&=&1+\frac{\psi^2}{L^2}(2\ell+a\psi)^2\cr
\Sigma^2&=&(r^2+\ell^2)^2\Delta_\psi-\psi^2(a\psi+2\ell)^2\Delta_r\cr
\omega&=&\frac{\Delta_r\psi(a\psi+2\ell)+a(r^2+\ell^2)\Delta_\psi}{\Sigma^2}.
\ea
Here $L$ is the asymptotic curvature radius and, as in the QKMV$_0$ metrics, the parameter $a$ corresponds to the angular momentum per unit mass. (However, for reasons explained in \cite{kn:shear}, $a$ is always \emph{negative} here.) The parameters  $M^*$ and $Q^*$ have slightly different physical interpretations from their QKMV$_0$ counterparts: for example, the charge density at the horizon is $Q^*/(r^2_h+\ell^2)$ rather than $Q^*/r^2_h$.

The asymptotic angular velocity in this case is given by
\begin{equation}\label{PHI}
\omega_{\infty} = \psi (a\psi + 2\ell)/L^2.
\end{equation}
Because $a$ is negative and $\ell$ positive, the velocity profile at infinity here is quite different from that of the QKMV$_0$ spacetime. When the boundary metric is expressed as in equation (\ref{DELTA}), the profile takes the form shown in Figure 2 (the functional form being that of a certain Weierstrass $\wp$-function).
\begin{figure}[!h]
\centering
\includegraphics[width=0.55\textwidth]{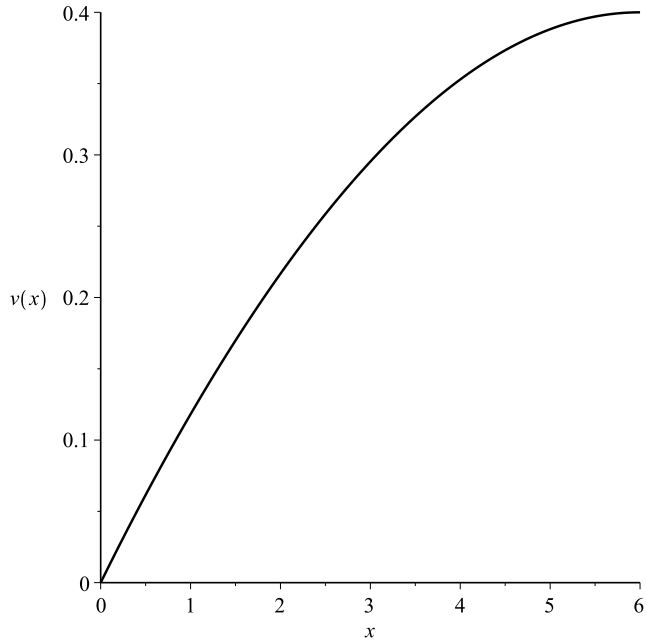}
\caption{$\ell$QKMV$_0$ velocity profile.}
\end{figure}

This is a suitable profile for the region far from the axis. A more realistic profile (similar to the ones in \cite{kn:huang}) can then be constructed by combining the lower part of Figure 1 with the upper part of Figure 2; see \cite{kn:shear} for this procedure. Of course, we do not claim that these particular functions do anything more than give a qualitative representation of the actual flow, but the overall shape is reasonable.

The electromagnetic potential is in this case
\be
\label{XI}
A\;=\;-\frac{Q^*r}{\rho^2}\dif
t+\frac{Q^*r\psi(a\psi+2\ell)}{\rho^2}\dif\zeta.
\ee

As before, we now construct the equations relating the boundary parameters to those of the bulk. It turns out that $\ell$ is dual to a length scale fixed by combining the parameter $a$ with the maximal dimensionless velocity $V$ of the plasma, near the boundary of the flow (so that, for example, $V \approx 0.4$ in Figure 2). Specifically, we have, from \cite{kn:shear},
\begin{equation}\label{PSI}
\ell^2\;=\;V|a|L.
\end{equation}
The equations which take the place of equations (\ref{VARDELTA}),(\ref{VARVARDELTA}), and (\ref{VARVARVARDELTA}) above are (see again \cite{kn:shear}),
\begin{equation}\label{OMEGA}
\varepsilon \;=\;M^*/(r^2_h + \ell^2),
\end{equation}
\begin{equation}\label{ALEPH}
T \;=\;{r_h(r_h^2+\ell^2)/L^2-2\pi M^* \over \pi (r_h^2+\ell^2)},
\end{equation}
\begin{equation}\label{BETH}
\mu = {Q^*r_h \over L[r_h^2+\ell^2+|a|L]}.
\end{equation}

Combining equations (\ref{OMEGA}) and (\ref{ALEPH}), one obtains exactly the same relation, equation (\ref{PI}), between $r_h$ and the temperature and energy density; so, once again, $r_h$ does not depend on the angular momentum once those parameters are fixed. The relation replacing equation (\ref{VARVARSIGMA}) is
\begin{equation}\label{GIMEL}
\tilde{\mu}\;=\;\sqrt{{\Big[8\pi \tilde{\varepsilon} \tilde{r}_h(\tilde{r}^2_{h}+V|\tilde{a}|) \; - \;(\tilde{r}^2_{h}+V|\tilde{a}|)^2 \; -\;\tilde{a}^2\Big]\,\tilde{r}_h^2\over 4\pi \Big[\tilde{r}^2_{h}+(1+V)|\tilde{a}|\Big]^2}}.
\end{equation}
As before, and as expected, this is a decreasing function of $\tilde{a}$: the effect of angular momentum is to lower the estimate of the chemical potential. As in the case of the $QKMV_0$ geometry, however, it is effectively constant for physical values of the parameters: if we take, as in \cite{kn:shear}, $V \approx 0.4$, and retain the same values for the other data as in the preceding section, then we find that the result is around 16600 MeV, the same value (to this level of approximation) as before. Once again, shearing angular momentum does reduce the predicted value of the chemical potential in principle, but not in practice, and so angular momentum is not helpful in this case either.

\addtocounter{section}{1}
\section* {\large{\textsf{4. Conclusion: Angular Momentum and More Realistic Holography}}}
Experimentally observed Quark-Gluon Plasmas are generically endowed with very large angular momenta, so it seems natural to incorporate this in holographic models. This can be done in a rather straightforward way, using frame-dragging. In this work we have seen that taking this step has an important side-benefit: it can, in the case in which the plasma rotates, usefully improve holographic estimates of the value of the quark chemical potential. Unfortunately, the same cannot be said in the case in which the angular momentum is carried by the internal shearing motion of the QGP. Thus, angular momentum is an important contributor to the project of rendering holography more realistic: but it is only part of the solution.

We saw that the rotational case differs so radically from the case of a shearing plasma because topologically spherical AdS black holes differ from their planar counterparts: the peculiarities of spherical topology force the angular momentum parameter $a$ to be bounded by the asymptotic AdS curvature radius $L$. When the angular momentum is large, one must generically expect that $a/L$ should be nearly unity, and this has a very strong distorting effect (see equations (\ref{C}), (\ref{D}), and (\ref{EGAD}) above) on the spacetime geometry, with no analogue in the shearing case. In some sense, rotating black holes are more sensitive to the asymptotic geometry than shearing black holes, and this geometric property manifests itself dually in the form of a greatly enhanced sensitivity of $\mu$ to the angular momentum in the rotating case.

However, there is another effect we are ignoring here, one which, like angular momentum, is in fact generically present in these collisions: \emph{strong magnetic fields}. These too have definite holographic representations, and one might well hope that, in combination with the effect discussed here, they too will help to give rise to more realistic holographic computations of the chemical potential. We will report on this issue elsewhere.

\addtocounter{section}{1}
\section*{\large{\textsf{Acknowledgement}}} The author thanks Prof. Soon Wanmei and Dr Jude McInnes for helpful discussions.

\end{document}